\documentclass[10pt]{amsart}

\usepackage{tikz}
\usepackage{amssymb,amsmath,latexsym,graphicx}
\usepackage{charter,eucal}
\usepackage{amssymb}
\usepackage{amscd, relsize,soul,tikz,color,xcolor}

\setul{}{1.1pt}

\usepackage[linktoc=page,colorlinks,linkcolor={red!80!black},citecolor={red!80!black},urlcolor={blue!80!black}]{hyperref}

\newtheorem{theorem}{Theorem }[section]
\newtheorem{lemma}[theorem]{Lemma}
\newtheorem{remark}[theorem]{Remark}
\newtheorem{corollary}[theorem]{Corollary}
\newtheorem{proposition}[theorem]{Proposition}
\newtheorem{question}[theorem]{\textsc{Question}}
\newtheorem{conjecture}[theorem]{\textsc{Conjecture}}
\newtheorem{definition}[theorem]{\textsc{Definition}}

\def\I{\mathrel{\mathrm{I}}}

\def\1{\mathrel{\mathbf{1}}}

\newcommand{\mQ}{\mathcal{Q}} 

\newcommand{\bI}{\mathrm{id}}

\newcommand{\id}{\mathrm{id}}
\newcommand{\C}{\mathbb{C}}
\newcommand{\F}{\mathbb{F}}

\newcommand{\eop}{\hspace*{\fill}$\blacksquare$}

\newcommand{\btt}{\begin{ttheorem}}
\newcommand{\ett}{\end{ttheorem}}

\newcommand{\bt}{\begin{theorem}}
\newcommand{\et}{\end{theorem}}
\newcommand{\bcc}{\begin{conjecture}}
\newcommand{\ecc}{\end{conjecture}}
\newcommand{\bc}{\begin{corollary}}
\newcommand{\bl}{\begin{lemma}}
\newcommand{\ec}{\end{corollary}}
\newcommand{\el}{\end{lemma}}
\newcommand{\bq}{\begin{question}}
\newcommand{\eq}{\end{question}}
\newcommand{\bp}{\begin{proposition}}
\newcommand{\ep}{\end{proposition}}
\newcommand{\br}{\begin{remark}}
\newcommand{\er}{\end{remark}}
\newcommand{\bd}{\begin{definition}}
\newcommand{\ed}{\end{definition}}

\newcommand{\U}{\ensuremath{\mathbf{U}}}

\newcommand{\PG}{\ensuremath{\mathbf{PG}}}

\newcommand{\Aut}{\ensuremath{\mathbf{Aut}}}

\newcommand{\mL}{\ensuremath{\mathcal{L}}}
\newcommand{\mH}{\ensuremath{\mathcal{H}}}

\newcommand{\mC}{\ensuremath{\mathcal{C}}}

\newcommand{\hP}{\ensuremath{\mathbf{P}}}

\newcommand{\A}{\ensuremath{\mathbf{A}}}
\newcommand{\mB}{\mathcal{B}}
\newcommand{\mP}{\mathcal{P}}

\newcommand{\hU}{\mathbf{U}}
\newcommand{\bP}{\mathbf{P}}
\newcommand{\R}{\mathbb{R}}

\author{Koen  Thas}

\address{Department of Mathematics,
Ghent University,
Krijgslaan 281, S25, B-9000 Ghent, Belgium}

\email{koen.thas@gmail.com}

\title[General Quantum Theory]{General Quantum Theory}

\date{October 2017}

\begin{document}

\maketitle

\begin{abstract}
Inspired by classical (``actual'') Quantum Theory over $\C$ and Modal Quantum Theory (MQT), which is a model of Quantum Theory over certain finite fields, we introduce General Quantum Theory as a Quantum Theory 
| in the K\o benhavn interpretation | over general division rings with involution, in which the inner product ``is'' a $(\sigma,1)$-Hermitian form $\varphi$. This unites all known such approaches in one and the same 
theory, and we show that many of the known results such as no-cloning, no-deleting, quantum teleportation and super-dense quantum coding, which are known in classical Quantum Theory over $\C$ and 
in some MQTs, hold for any General Quantum Theory. On the other hand, in many General Quantum Theories, a geometrical object
which we call ``quantum kernel'' arises, which is invariant under the unitary group $\hU(V,\varphi)$, and which carries the geometry of a so-called polar space. This object cannot be seen in classical Quantum Theory over $\C$, but it is present, for instance, in all know MQTs.  We use this object to construct new quantum (teleportation) coding schemes, which mix quantum theory with the geometry of the quantum kernel (and the action of the unitary group). 
We also show that in characteristic $0$, every General Quantum Theory over an algebraically closed field behaves like classical Quantum Theory over $\C$ at many levels, and that all such theories share one model, which we pin down
as the ``minimal model,'' which is countable and defined over $\overline{\mathbb{Q}}$. Moreover, to make the analogy with classical Quantum Theory even more striking, we show that Born's rule holds in any such theory. So all such theories are not modal at all. Finally, we obtain an extension theory for General Quantum Theories in characteristic $0$ which allows one to extend any such theory over algebraically closed fields (such as classical complex Quantum Theory) to larger theories in which a quantum kernel is present. In this sense, these singular objects are always virtually around in abundance.
\\
\end{abstract}



 \setcounter{tocdepth}{1}
\tableofcontents

\medskip
\section{Introduction}
\label{}

{\footnotesize
\begin{flushright}
``One can give good reasons why reality cannot at all be represented by a continuous field.\\ 
From the quantum phenomena it appears to follow with certainty that a finite system of finite energy\\ 
can be completely described by a finite set of numbers (quantum numbers).''\\ 
(A. Einstein, from ``The Meaning of Relativity'') \\
\end{flushright}
}

\bigskip
In classical Quantum Theory following the K\o benhavn interpretation | in some papers called ``Actual Quantum Theory'' (AQT) | the state space is a Hilbert space (foreseen with the standard inner product). More precisely:
\begin{itemize}
\item[(*)]
a physical quantum system is represented by a Hilbert space $\mH = \Big((\mathbb{C}^\omega,+,\cdot),\langle \cdot,\cdot \rangle\Big)$, with $\langle \cdot,\cdot \rangle$ the standard inner product and $\omega$ allowed to be non-finite;
\item[(*)]
the standard inner product $\langle \cdot,\cdot \rangle$ sends $\Big((x_1,\ldots,x_{\omega}),(y_1,\ldots,y_{\omega})\Big)$ to $\overline{x_1}y_1 + \cdots + \overline{x_{\omega}}y_{\omega}$ (or $x_1\overline{y_1} + \cdots + x_{\omega}\overline{y_{\omega}}$), where $\overline{c}$ is the complex conjugate of $c \in \mathbb{C}$; complex conjugation is an involutary automorphism of the field $\mathbb{C}$; 
\item[(*)]
up to complex scalars, pure states (wave functions) are represented by nonzero vectors in $\mathbb{C}^\omega$; usually, one considers normalized vectors; 
\item[(*)]
observables are represented by linear operators of $\mathbb{C}^\omega$ that preserve $\langle \cdot,\cdot \rangle$, that is, {\em unitary operators}. If $\omega$ is finite, unitary operators correspond to nonsingular complex $(\omega \times \omega)$-matrices $U$ such that $UU^* = \id$; 
\item[(*)]
measuring an observable $A$ in a system described by the wave function $\vert \psi \rangle$, amounts to collapsing $\vert \psi \rangle$ into one of the orthogonal eigenvectors $\vert \psi_i \rangle$ of $A$, yielding as measurement the corresponding eigenvalue $\lambda_i$; 
\item[(*)]
composite product states correspond to tensor products $\vert \psi_1 \rangle \otimes \vert \psi_2 \rangle \in \mH_1 \otimes \mH_2$; if a state in $\mH_1 \otimes \mH_2$ 
is not a product state, it is entangled;
\item[(*)]
one follows Born's rule, which says that $\vert \langle \psi, \psi_i \rangle \vert^2$ is the probability that the measurement $\lambda_i$ will be made. 
\end{itemize}

In \cite{MQT} the authors propose to consider ``Modal Quantum Theory'' (MQT) as a toy model for AQT, in which $\mathbb{C}$ is replaced by
a finite field $\mathbb{F}_q$. Since an innner product is not defined on vector spaces over a finite field, the authors drop this aspect of the theory, but still build an interesting theory, obtaining for instance a no-cloning result, see \cite{MQT}. The authors claim that by dropping inner products or variations, one does not have notions such as ``orthogonality'' at hand, and that can't be what one wants. Also, one does not have probabilities at one's disposal in the same way as in AQT. In \cite{MQT} the authors focus primarily on the case $\mathbb{F}_q = \F_2$. 
In other papers such as \cite{Lev}, the authors consider vector spaces over finite prime fields $\mathbb{F}_p$ with the property that $-1$ is not a square in $\mathbb{F}_p$, but it is in $\mathbb{F}_{p^2}$, so as to have, besides the similarity between the fields $\mathbb{R}$ and $\mathbb{C}$, a Hermitian bilinear form at hand which shares many important aspects with the inner product $\langle \cdot,\cdot \rangle$. For one, the corresponding natural operators in that setting are also unitary operators, and now orthogonal vectors are well defined.\\

In various other parts of literature, allusions have been made on Quantum Theory over other rings (commutative or not), but mostly the discussions are speculative, and in any case not detailed. In this paper, we wish to address both MQT and variations over other rings, and formalize both ideas into one theory, namely Quantum Theory over division rings. \\

We introduce General Quantum Theory (GQT) as a Quantum Theory 
in the K\o benhavn interpretation over general division rings with involution, in which the inner product is replaced by a $(\sigma,1)$-Hermitian form $\varphi$. This unites all known such approaches in one and the same 
theory, and we show that many of the known results such as: 
\begin{itemize}
\item[$\star$]
no-cloning (a result first obtained for AQT by Wootters and Zurek \cite{WZclone,WZclone2} and Dieks \cite{Dieks} in 1982, and later adapted to MQT over some finite fields in \cite{MQT}),
\item[$\star$]
no-deleting (obtained by Pati and Braunstein in \cite{nodelete} in 2000), 
\item[$\star$]
quantum teleportation (obtained in AQT by Bennet, Brassard, Cr\'{e}peau, Josza and Wootters in \cite{QT} in 1993),  and 
\item[$\star$]
super-dense quantum coding (described for AQT in Bennet and Wiesner \cite{Bennet} in 1992), 
\end{itemize}
hold for any General Quantum Theory. One has to be a bit more careful in the proofs since multiplication is not necessarily commutative anymore, and often we have to make different proofs when the characteristic is $2$. \\

On the other hand, in some General Quantum Theories, a geometrical object
which we call ``quantum kernel'' arises, which is invariant under the unitary group $\hU(V,\varphi)$, and which carries the geometry of a so-called ``polar space.'' 
This is a combinatorial object which has a long history of study, and which is, if not trivial, the natural geometric module on which the unitary group acts. 
This object cannot be seen in classical Quantum Theory over $\C$, but it is present, for instance, in all know MQTs (but also many others over algebraically closed fields or other division rings).  We use this object to construct new quantum (teleportation) coding schemes, which mix quantum theory with the geometry of the quantum kernel (and the action of the unitary group). It is precisely this object (over finite fields) which has been used over and over again in Quantum Theory to construct and understand, e.g., maximal sets of mutually unbiased bases \cite{KTPauli,KTUMUB}.\\

We observe that in MQT there is no need to only consider prime fields with the aforementioned arithmetic property, and provide an elegant model in which one can see all MQTs in a fixed characteristic at once. 
We also show that in characteristic $0$, every General Quantum Theory over an algebraically closed field behaves at many levels as classical Quantum Theory over the complex numbers, and that all such theories share one model, which we pin down
as the ``minimal model,'' which is countable and defined over $\overline{\mathbb{Q}}$. Moreover, to make the analogy with AQT even more striking, we show that Born's rule holds in any such theory. (So these theories are not modal at all.) \\

Because of these results, we argue that one can replace classical Quantum Theory over $\C$ by other Quantum Theories which carry the same information, but come with extra geometric tools, or by finite Quantum Theories (in which the quantum kernel is very well understood), or in the minimal model, which is countable. On the other hand, we hope to provide a deeper insight in these theories by introducing the present unification. Also, it is important to mention Wootters's paper \cite{Wootfg}, in which the underlying finite geometrical nature
of several fundamental quantum theoretical problems is investigated. The existence of the quantum kernel certainly adds a foundational aspect to the general theory.


\medskip
\section{Modal quantum theory and variations in the general setting}
\label{secMQT}

In \cite{MQT} the authors introduce {\em Modal Quantum Theory} (MQT) as a finite model for AQT, in which $\mathbb{C}$ is replaced by
a finite field $\mathbb{F}_q$. Innner products are not defined on vector spaces over a finite field, and hence the authors ignore this aspect of the theory. As such, one cannot speak of ``orthogonal states.''
In  \cite{Lev}, the authors consider vector spaces over finite prime fields $\mathbb{F}_p$ with the property that $-1$ is not a square in $\mathbb{F}_p$, but it is in $\mathbb{F}_{p^2}$, so as to have, besides the similarity between the fields $\mathbb{R}$ and $\mathbb{C}$, a Hermitian bilinear form at hand which shares many important aspects with the inner product $\langle \cdot,\cdot \rangle$. 

There is no need for restricting the theory to primes with the aforementioned property, as we will see below.

Let $q$ be any prime power; then up to isomorphism $\mathbb{F}_q$ has a unique extension of degree $2$, namely $\mathbb{F}_{q^2}$. The map
\begin{equation}
\gamma: \mathbb{F}_{q^2} \mapsto \mathbb{F}_{q^2}: a \mapsto a^q
\end{equation}
sends each element of $\mathbb{F}_q$ to itself, while being an involutory automorphism of $\mathbb{F}_{q^2}$ (that is, $u^{\gamma^2} = u$
for each $u \in \mathbb{F}_{q^2}$). 

Let $n \in \mathbb{N}$ be any positive integer different from $0$; then if $V = V(n,q^2)$ is the $n$-dimensional vector space over $\mathbb{F}_{q^2}$, define for $x = (x_1,\ldots,x_n)$ and $y = (y_1,\ldots,y_n)$ in $V$, 
\begin{equation}
\Big\langle x,y \Big\rangle := x_1^{\gamma}y_1 + \cdots + x_n^{\gamma}y_n. 
\end{equation}

Then for $\rho, \rho' \in \mathbb{F}_{q^2}$ we have that 
\begin{equation}
\Big\langle \rho x,\rho' y\Big\rangle = \rho^{\gamma}\Big\langle x,y \Big\rangle \rho',\ \mbox{and}\ \Big\langle x,y\Big\rangle^{\gamma} = \Big\langle y,x\Big\rangle.
\end{equation}

The linear $(n \times n)$-matrices $U$ which preserve the form $\langle \cdot,\cdot \rangle$ precisely are unitary matrices: $(n \times n)$-matrices $U$ for which $U^* U$ is the $(n \times n)$-identity matrix, where $U^* := {(U^{\gamma})}^T$. \\

In this model of QT, $\mathbb{F}_{q^2}$ plays the role of $\mathbb{C}$, $\mathbb{F}_q$ the role of $\mathbb{R}$, $\gamma$ the role of complex conjugation, and $\langle \cdot,\cdot \rangle$ the role of inner product. No-cloning and no-deleting can be obtained in these MQTs, but we will handle this in a much more general context in sections \ref{noclone} and \ref{nodelete}. By choosing any element $\kappa$ in $\mathbb{F}_{q^2} \setminus \F_q$, we can represent each element of $\mathbb{F}_{q^2}$ uniquely as 
\begin{equation}
a + \kappa b,
\end{equation}
with $a, b \in \F_q$. So viewed from this representation, the situation at least looks ``a little classical.''

\medskip
\section{``All at once representation''}

Let $p$ be any prime, and let $\overline{\mathbb{F}_p}$ be an algebraic closure of $\mathbb{F}_p$ (such a field is uniquely determined up to isomorphism). For any power $q$ of $p$, one has that $\overline{\mathbb{F}_q} = \overline{\mathbb{F}_p}$. For any positive integer $m$, 
$\mathbb{F}_p$ contains {\em precisely one} subfield which is isomorphic to $\mathbb{F}_{p^m}$. So all finite degree extension fields of $\mathbb{F}_p$ are subfields of $\overline{\mathbb{F}_p}$. 

The automorphism group $\Aut(\overline{\mathbb{F}_p})$ of $\overline{\mathbb{F}_p}$ consists of all maps
\begin{equation}
\kappa_i:  \overline{\mathbb{F}_p} \mapsto \overline{\mathbb{F}_p}:\ \ x \mapsto x^{p^i},
\end{equation}
with $i$ any positive integer. So $\Aut(\overline{\mathbb{F}_p})$ is an infinite group which is isomorphic to the circle group. 

Using this information, we can represent all MQTs in a fixed characteristic $p$ as follows:\\

$\left\{
\begin{tabular}{p{.9\textwidth}}
\begin{itemize}
\item
take any element $\nu = \kappa_i \in \Aut(\overline{\mathbb{F}_p})$;
\item
let $p^i$ be the corresponding power;
\item
$\nu$ fixes the subfield $\mathbb{F}_{p^i}$ pointwise, and acts as an involutory automorphism of $\mathbb{F}_{p^{2i}}$;
\item
now, for any positive integer or infinite cardinal number $m$, define the MQT $\Big(V(m,\mathbb{F}_{p^{2i}}),\langle \cdot,\cdot \rangle_{\nu} \Big)$.
\end{itemize}
\end{tabular}\right.$\\

So we can represent all MQT-theories in the affine space $\A^3(\mathbb{Z})$, where each point in $\mathbb{Z}_{>0} \times \mathbb{Z}_{>0} \times P$, with $P$ the set of primes, 
defines an MQT.

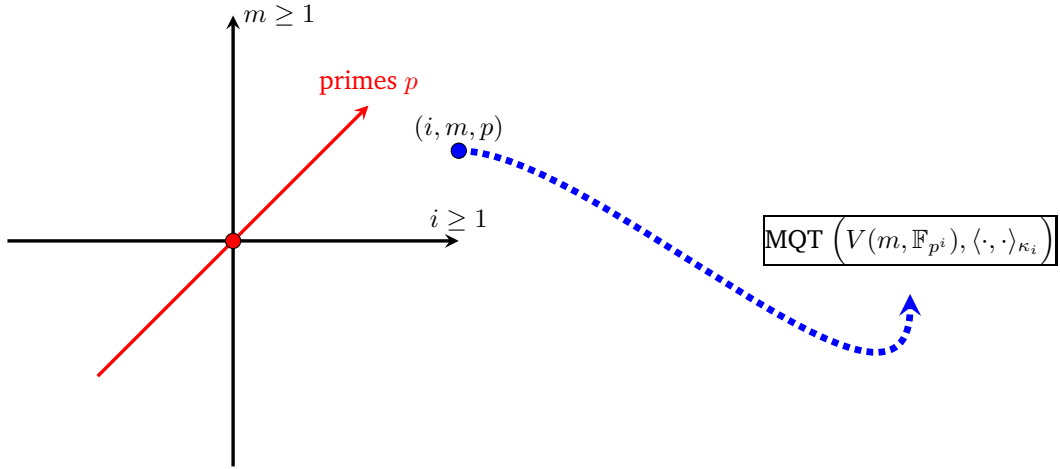
\begin{center}
\item
\begin{tikzpicture}[x=0.5cm,y=0.5cm,z=0.3cm,>=stealth,scale=1.2]
\draw[->] [line width=0.45mm, black ] (xyz cs:x=-5) -- (xyz cs:x=5) node[above] {$i \geq 1$};
\draw[->] [line width=0.45mm, black ] (xyz cs:y=-5) -- (xyz cs:y=5) node[right] {$m \geq 1$};
\draw[->] [line width=0.45mm, red ] (xyz cs:z=-5) -- (xyz cs:z=5) node[above] {primes $p$};

\draw[fill=red] (0,0) circle (.5ex) node[above] {};
\draw[fill=blue] (5,2) circle (.5ex) node[above] {$(i,m,p)$};
\draw[->,help lines,shorten >=3pt] [line width=0.9mm, blue,dotted ] (5,2) .. controls (8,2) and (15,-5) .. (15,-1);
\node[circle] at (15,0) {\frame{MQT $\Big(V(m,\F_{p^i}),\langle \cdot,\cdot \rangle_{\kappa_i} \Big)$}};
\end{tikzpicture}
\item
``Universal MQT-space:'' with each point in the first quadrant corresponds an MQT-theory.
\end{center}

\medskip
\section{Why division rings?}

\subsection{Projective geometries over division rings}

Let $D$ be a division ring, let $n \in \mathbb{N}$, and let $V = V(n,D)$ be the $n$-dimensional (left or right) vector space over $D$. We define the {\em $(n - 1)$-dimensional (left or right) projective space} $\PG(n - 1,D)$  as being the geometry of vector subspaces of $V$. For example: the points of $\PG(n - 1,D)$ correspond to the vector lines of $V(n,D)$, and 
the lines of $\PG(n - 1,D)$ correspond to the vector planes of $V(n,D)$. A point of $\PG(n - 1,D)$ lies on a line of $\PG(n - 1,D)$ if the corresponding vector line lies in the corresponding vector plane. And so on.  The choice of ``left'' or ``right'' does not affect the isomorphism class of the space.

If $D = \F_q$ is the finite field with $q$ elements ($q$ a prime power), one  also writes $\PG(n - 1,q)$ instead of 
$\PG(n - 1,\F_q)$. 

Geometrical properties of a $\PG(n - 1,D)$ depend on the choice of division ring (see \cite{Ernie} for much more).

\subsection{Veblen-Young result}

In AQT, each quantum state is represented by an element $\vert \psi \rangle \ne \overline{0}$ of $\mathbb{C}^n$ {\em up to a scalar}, that is,
$\vert \psi \rangle$ and $c\cdot \vert\psi\rangle$ represent the same state ($c \ne 0$). So the geometry of quantum states in $\mathbb{C}^n$ is in fact the geometry of its rays, that is, the corresponding projective geometry $\PG(n - 1,\mathbb{C})$. 

There is a notion of {\em axiomatic projective space}, which is defined to be a  geometry of points and lines which satisfies certain axioms, of which the following are the essential ones:
\begin{itemize}
\item
Two arbitrary different points are incident with precisely one line.
\item
Given two different intersecting lines $U, V$ with $U \cap V = \{w\}$, and given two different lines $X, Y$ meeting both $U$ and $V$, and in points different from $w$, 
it holds that $X$ and $Y$ also meet in precisely one point.
\end{itemize}

Using only these axioms, one naturally defines a notion of {\em linear subspace}, and then one defines the dimension of the geometry in a natural way as well.  

A remarkable result of Veblen and Young \cite{VY} states  that if the dimension $n - 1$ of such a space is at least three, it is isomorphic to some $\PG(n - 1,D)$ with $D$ a division ring. This is not true when the dimension is less than three. \\

\subsection{Setting}

If we imagine that each ``general Hilbert space'' should include a module over some ring (as a generalization of vector space over $\C$) which serves as the state space, and if we agree that the corresponding space  of rays should be at least an axiomatic projective geometry, then Veblen-Young's theory leads us to \ul{vector spaces over division rings}, and this is our general setting. \\

In the next section, we introduce the most natural candidates for replacing the inner product we need to set up the theory.

\medskip
\section{$(\sigma,1)$-Hermitian forms}
\label{Herm}

Let $k$ be a division ring. An {\em anti-automorphism} of $k$ is a map $\gamma: k \mapsto k$ such that
\begin{itemize}
\item
$\gamma$ is bijective;
\item
for any $u, v \in k$, we have $\gamma(u + v) = \gamma(u) + \gamma(v)$;
\item
for any $a, b \in k$, we have $\gamma(ab) = \gamma(b)\gamma(a)$.
\end{itemize}

\subsection{Examples}

\begin{itemize}
\item
If $k$ is a commutative field, then anti-automorphisms and automorphisms coincide. \\ 
\item
Let $\mathbb{H}$ be the quaternions, i.e., the set $\{ a + bi + cj + dk\ \vert\ a, b, c, d \in \mathbb{R} \}$ with symbols $i, j, k$ satisfying 
$i^2 = j^2 = k^2 = ijk = -1$. Then 
\begin{equation}
\gamma: \mathbb{H} \mapsto \mathbb{H}: a + bi + cj + dk \mapsto a - bi - cj - dk 
\end{equation}
is an anti-automorphism. Moreover, the restriction to the subfield $\C$ (generated by $1$ and $i$) precisely is complex conjugation. 
\item
The fields $\mathbb{Q}$ and $\mathbb{R}$ do not admit nontrivial automorphisms. 
\end{itemize}

\subsection{Hermitian forms}

Suppose that $k$ is a division ring, and suppose $\sigma$ is an anti-automorphism of $k$. Let $V$ be a right vector space over $k$.
A {\em $\sigma$-sesquilinear form} on $V$ is a map $\nu: V \times V \mapsto k$ for which we have the following properties: 
\begin{itemize}
\item
for all $a,b,c,d \in V$ we have that $\nu(a + b,c + d) = \nu(a,c) + \nu(b,c) + \nu(a,d) + \nu(b,d)$;
\item
for all $a, b \in V$ and $\alpha, \beta \in k$, we have that $\nu(a\alpha,b\beta) = \sigma(\alpha)\nu(a,b)\beta$.
\end{itemize}

We have that $\nu$ is reflexive if and only if  there exists an $\epsilon \in k$ such that for all $a, b \in V$, we have
\begin{equation}
\nu(b,a) = \sigma\Big(\nu(a,b)\Big)\epsilon.
\end{equation}

Such sesquilinear forms are called {\em $(\sigma,\epsilon)$-Hermitian}. If $\epsilon = 1$ and $\sigma^2 = \id \ne \sigma$, then we speak of 
a Hermitian form. 

Clearly, \ul{the standard inner product in a classical Hilbert space over $\C$ is a Hermitian form.}  In fact, \ul{{\em any} inner product on a complex Hilbert space
is a Hermitian form.} \\

\subsection{Standard $(\sigma,1)$-Hermitian forms}

If $k$ is a division ring with involution $\sigma$, the {\em standard $(\sigma,1)$-Hermitian form} on the right vector space $V(d,k)$, is given by

\begin{equation}
\Big\langle x,y \Big\rangle\ :=\ x_1^\sigma y_1 + \cdots + x_d^\sigma y_d, 
\end{equation}
where $x = (x_1,\ldots,x_d)$ and $y = (y_1,\ldots,y_d)$. 

In the case that $\sigma = \id$, we obtain a form which is usually called {\em symmetric}; it is not a proper Hermitian form, but still comes in handy in some situations
(for example in cases of field reduction: ``real Hilbert spaces'' have often been considered in Quantum Theory; see e.g. \cite{Wootreal,Wootreal2,Wootreal3}).

\subsection{Morphisms of $(\sigma,1)$-Hermitian forms}
\label{morph}

An {\em automorphism} of a $(\sigma,1)$-Hermitian form $\varphi$ on the $k$-vector space $V$, is a bijective linear operator $\omega: V \mapsto V$
which preserves $\varphi$, that is, for which 
\begin{equation}
\varphi(\omega(x),\omega(y)) = \varphi(x,y)
\end{equation}
for all $(x,y) \in V \times V$. The group of all such automorphisms is called the {\em unitary group}, and denoted $\U(V,\varphi)$. 

\subsection*{Example}

Let $k = \C$, $\sigma$ be complex conjugation, and $V = V(n,\C)$. Then $\hU(V,\varphi) = \mathbf{GU}_n(\C) = \mathbf{U}(n)$.

\medskip
\section{GQT}

Throughout this paper, if we speak of ``division ring with involution,'' we mean a division ring with an involutory anti-automorphism.\\

$\left\{
\begin{tabular}{p{.9\textwidth}}
From now on, we propose to depict a physical quantum system by a \ul{general Hilbert space $\mH = \Big((V(\omega,k),+,\cdot),\langle \cdot,\cdot \rangle\Big)$, with $k$ a division ring with involution $\sigma$, and $\Big\langle \cdot,\cdot \Big\rangle$ a $(\sigma,1)$-Hermitian form.} \\
\end{tabular}
\right\}$

\bigskip
If we speak of ``standard GQT,'' we mean that given $\sigma$, the general Hilbert space comes with the standard $(\sigma,1)$-Hermitian form. Also, as some fields such as 
the reals and the rational numbers do not admit nontrivial involutions, they only can describe ``improper'' quantum systems. By extension of Quantum Theories (cf. 
section \ref{ext}), this is no problem (as often has been the case when switching between AQT over $\C$ and $\R$).

\medskip
\section{Matrix representation and mixed Hilbert spaces}

Consider general Hilbert spaces $\mH_1$ and $\mH_2$, with respective $(\sigma,1)$-Hermitian forms $f_1$ and $f_2$ defined over the same division ring $D$ with involution $\sigma$. In the same way as for 
bilinear forms, we can represent $f_1$ and $f_2$ by using matrices when both vector spaces are finite-dimensional. Let the matrices be $A_1$ and $A_2$ respectively; then these 
are Hermitian matrices.
If $x, y \in \mH_1$ and $u, v \in \mH_2$, then
\begin{equation}
f_1(x,y) = xA_1\sigma(y^T)\ \ \mbox{and}\ \ f_2(u,v) = uA_2\sigma(v^T).
\end{equation}

One easily verifies that 
\begin{equation}
(x \otimes y)\Big(A_1 \otimes A_2\Big) (\sigma((y \otimes v)^T)) = (xA_1\sigma(y^T))\otimes(uA_2\sigma(v^T)),
\end{equation}
and moreover, $\sigma((A_1 \otimes A_2)^T) = \sigma(A_1^T) \otimes \sigma(A_2^T) = A_1 \otimes A_2$, so $A_1 \otimes A_2$ is Hermitian as well. 
By linear expansion, it follows that $f_1 \otimes f_2$ defines a $(\sigma,1)$-Hermitian form over $D$ on the tensor product $\mH_1 \otimes \mH_2$.  Similar 
reasoning can be done in the infinite-dimensional case. \\

In conclusion, we can say that in every GQT, composite systems are represented in tensor products of the (generalized) Hilbert spaces (with $(\sigma,1)$-forms), as in AQT. 
Note that the above matrix formalism is not violated if one considers {\em different} $(\sigma,1)$-Hermitian forms in $\mH_A$ and $\mH_B$. In such a way, one can 
construct {\em mixed composite systems} in a strong sense of the word, exposing mixed properties of the Hermitian forms from the components. 

\medskip
\subsection{Example: AQT}

If we put $k = \C$ and take the standard inner products on $\mH_1$ and $\mH_2$, then $A_1$ and $A_2$ are identity matrices, and so is $A_1 \otimes A_2$, 
so in that case $f_1 \otimes f_2$ is the standard inner product on $\mH_1 \otimes \mH_2$.

\medskip
\subsection{Standard GQT}

Given any division ring $D$ with involution $\sigma$, the standard $(\sigma,1)$-Hermitian form (with respect to a chosen basis) gives rise to the identity matrix.

\medskip
\section{The quantum kernel}

Let $k$ be any division ring with involution $\gamma$, and let $\Big\langle \cdot,\cdot \Big\rangle_{\gamma}$ be a  $(\sigma,1)$-Hermitian form on the right $k$-vector space $V$. To these data there is associated a geometrical object which we will describe below, and which we will call the {\em quantum kernel} relative to the given data. This geometric object is \ul{not visible over $\C$}. Also, it essentially consists of the vectors which are orthogonal to themselves, and thus they cannot be normalized. This object was not described in \cite{MQT}, despite its truly different nature than its complement in the modal Hilbert space.

\medskip
\subsection{Definition}

For any $v \in V$, define
\begin{equation}
\pi(v) := \{ w \in V \ \vert\ \  \Big\langle v,w \Big\rangle_{\gamma} = 0 \}.
\end{equation}

Note that $\Big\langle \cdot,\cdot \Big\rangle_{\gamma}$ is reflexive. Further, for any vector subspace $W$ of $V$, define
\begin{equation}
\pi(W) := \bigcap_{w \in W}\pi(w).
\end{equation}

On the associated projective space $\hP(V)$, we obtain a map $\pi$, denoted in the same way, which maps points to hyperplanes and hyperplanes to points. Moreover, we have that $\pi^2(v) = v$ for any vector $v \in V$. By definition, this means that $\pi$ is a {\em polarity} of $V$, or $\hP(V)$. 
We associate the following geometric structure, called {\em polar space} and denoted $\mP(V,\pi)$, to $(\hP(V),\pi)$ (and refer to the books \cite{Ernie,Uber} for all the details):
\begin{itemize}
\item
The points of $\mP(V,\pi)$ are those points corresponding to self-orthogonal vectors $v$ in $V$ (so for which $\Big\langle v,v \Big\rangle_{\gamma} = 0$). 
\item
The subspaces of $\mP(V,\pi)$ are those projective subspaces $\alpha$ of $\hP(V)$ such that $\alpha \subseteq \pi(\alpha)$; so, if $W$ is the vector subspace of $V$ corresponding to $\alpha$, then $W \subseteq \cap_{w \in W}\pi(w)$. 
\end{itemize}

We now investigate some basic properties of $\mQ := \mP(V,\pi)$.

\subsubsection{Structure of $\pi(v)$}

To fix ideas, we let $\Big\langle \cdot,\cdot \Big\rangle_{\gamma}$ be the standard $(\gamma,1)$-Hermitian form; everything in this paragraph is easily adaptable to general
$(\gamma,1)$-Hermitian forms.

Let $v = (v_1,\ldots,v_d)$; then the vectors $w = (w_1,\ldots,w_d)$ of $\pi(v)$ are the solutions of $v_1x_1^{\gamma} + \cdots + v_dx_d^{\gamma} = 0$, so after letting $\gamma$ act on both sides of the equation, they are the solutions of 
\begin{equation}
x_1v_1^{\gamma} + \cdots + x_dv_d^{\gamma} = 0.
\end{equation}

If $v$ is not the zero vector, this means that $\pi(v)$ is a hyperplane in $V$ (= a vector subspace of dimension $d - 1$).

\subsubsection{Polar inequality}
\label{polar}

Let $A$ and $B$ be subsets of $V$ such that $A \subseteq B$; then obviously $\pi(B) \subseteq \pi(A)$. In particular, if $A$ and $B$ are 
subspaces of $V$ and $B$ is absolute, then 
\begin{equation}
A \subseteq B \subseteq \pi(B) \subseteq \pi(A),
\end{equation}
so  that $A$ is absolute as well.

\subsubsection{Collinearity}

Let $\mP(V,\pi)$ be as above. We say that points $x, y$ of $\mP(V,\pi)$ are {\em collinear} if they verify $\Big\langle x,y \Big\rangle = 0$. This is equivalent to 
saying that such points are collinear if there is some line of $\mP(V,\pi)$ which contains both $x$ and $y$.

\subsubsection{Self-orthogonality}

By the previous subsection, we have that if $B$ is absolute, then all its vectors are also absolute, that is, for each $b \in B$ we have 
that $b \in \pi(b)$. In particular, $\langle b \vert b \rangle = 0$, so that $b$ is self-orthogonal. The following observation is a direct corollary:

\bp
$\mP(V,\pi)$ is empty if and only if there are no nonzero self-orthogonal vectors.
\ep



In general Modal Quantum Theory, the situation is very different, and hence the quantum kernel carries extra information about the quantum system. We review some important examples in detail, starting with complex AQT.

\medskip
\subsection{Example: the classical Hilbert space}

In the classical AQT-setting, so with $V \cong \mathbb{C}^d$,  $\gamma$ complex conjugation and $\Big\langle \cdot,\cdot \Big\rangle$ standard, the corresponding Hermitian form is also an inner product, so there are no non-trivial self-orthogonal vectors. 

\bp
The quantum kernel for AQT is the empty space.
\ep

\medskip
\subsection{Example: general modal Quantum Theory}
\label{h3q}

Now we pass to finite fields. We use the setting of section \ref{secMQT}. In that case, the Hermitian form is not an inner product, and the quantum kernel has a highly nontrivial structure.
We consider $k = \F_{q^2}$, $V = V(4,q^2)$, and $\pi$ as above; the involution $\gamma$ is given by $\gamma: \F_{q^2} \mapsto \F_{q^2}: r \mapsto r^q$. Consider the polar space
$\mQ = \mP(V,\pi)$ in $\PG(3,q^2)$; once chosen suitable homogeneous coordinates, its points satisfy the equation
\begin{equation}
X_0^{q + 1} + X_1^{q + 1} + X_2^{q + 1} + X_3^{q + 1} = 0.
\end{equation} 

We list some properties (all of which can be found and proved in \cite[chapter 3]{FGQ2}):
\begin{itemize}
\item[(1)]
each point of $\mQ$ is contained in $q + 1$ lines; there are $(q + 1)(q^3 + 1)$ points;
\item[(2)]
each line of $\mQ$ contains $q^2 + 1$ points; there are $(q^2 + 1)(q^3 + 1)$ lines;
\item[(3)]
given a point $x$ of $\mQ$ and a line $U$ of $\mQ$ that does not contain $x$, there is a unique line $V$ which contains $x$
and meets $U$ (in one point). 
\end{itemize}

Since $\mQ$ only has points and lines as full linear subspaces, one says it has {\em rank $2$}. Because of properties (1)-(2)-(3), it is a {\em generalized quadrangle} of order $(q^2,q)$, and usually it is denoted as $\mH(3,q^2)$. 

 We will use this very important example to construct quantum coding schemes in section \ref{codes}. \\

\medskip
\subsection{All-in-one axiom}

Now let $k$ be an arbitrary division ring with involution $\gamma$, and let $\Big\langle \cdot,\cdot \Big\rangle_{\gamma}$ be a $(\gamma,1)$-Hermitian form on the right $k$-vector space $V$ (which we suppose to be of finite dimension, for the sake of convenience). In this section, we suppose that $\mQ \ne \emptyset$. 
Let $\mP$ be the point set of $\mQ$, and let $\mL$ be the line set. The following defining axiom can be verified \cite{Uber}:

\begin{itemize}
\item[{\bf One-or-All} (OoA)]
If $x \in \mP$ and $U \in \mL$, and $U$ does not contain $x$, then either all points of $U$ are collinear with $x$, or precisely one point is. 
\end{itemize}

Note that property (3) for the case $\mQ = \mH(3,q^2)$ is a special case of the ``One-or-All'' axiom. A spectacular result of Buekenhout and Shult \cite{BueShu} characterizes
polar spaces as those point-line spaces that satisfy (OoA).

\bt[Buekenhout and Shult \cite{BueShu}]
Let $\Gamma = (\mP,\mL,\I)$ be a point-line geometry which satisfies (OoA). Then there exists a division ring $k$ with involution $\gamma$, 
a right $k$-vector space $V$ and a $(\gamma,1)$-Hermitian form, such that $\Gamma = \mP(V,\pi)$. 
\et

\begin{center}
\item
\begin{tikzpicture}[scale=1]
\draw [line width=0.45mm, black ] (-4,0) -- (0,0) node[below] {$Y$};
\draw [line width=0.45mm, black ] (2,0) -- (6,0) node[below] {$Y$};

\draw [line width=0.45mm, black ] (-2,2) -- (-2,0) node[above] {};

\draw [line width=0.45mm, black ] (4,2) -- (4,0) node[above] {};
\draw [line width=0.45mm, black ] (4,2) -- (3,0) node[above] {};
\draw [line width=0.45mm, black ] (4,2) -- (5,0) node[above] {};
\draw [line width=0.45mm, black ] (4,2) -- (3.5,0) node[above] {};
\draw [line width=0.45mm, black ] (4,2) -- (4.5,0) node[above] {};
\draw[fill=blue] (-2,2) circle (.5ex) node[right] {$x$};
\draw[fill=blue] (4,2) circle (.5ex) node[right] {$x$};
\draw[fill=blue] (-2,0) circle (.5ex) node[above] {};
\draw[fill=blue] (4,0) circle (.5ex) node[above] {};

\draw[fill=blue] (3,0) circle (.5ex) node[above] {};
\draw[fill=blue] (3.5,0) circle (.5ex) node[above] {};
\draw[fill=blue] (4.5,0) circle (.5ex) node[above] {};
\draw[fill=blue] (5,0) circle (.5ex) node[above] {};

\end{tikzpicture}
\item
\item
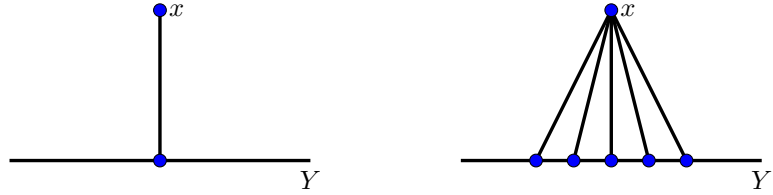
Buekenhout-Shult One-or-All axiom.
\end{center}


\medskip
\subsection{Automorphisms of $V \setminus \mQ$}
\label{autq}

In AQT, the natural transformations to work with are unitary transformations; they are precisely the linear operators which preserve the inner product. 
in GQT, one would like to have these transformations 
at one's disposal as well, and this indeed remains the fact in the general theory. The fact that $\mQ \ne \emptyset$ is no problem: in fact, 
it is the main geometric module to understand unitary groups in the first place.

By definition of the quantum kernel, any linear operator $\omega$ which preserves $\Big \langle \cdot,\cdot \Big\rangle$, also preserves self-orthogonality: if $\Big \langle x,x \Big\rangle = 0$, then
\begin{equation}
\Big \langle \omega(x),\omega(x) \Big\rangle = \Big \langle x,x \Big\rangle = 0.
\end{equation}

So $\hU(V,\Big \langle \cdot,\cdot \Big\rangle)$ acts on $\mQ$, and on $V \setminus \mQ$. The action on the polar space $\mQ$ is highly nontrivial, and well known. 
For instance, when $D = \F_{q^2}$ and $d = 3$ or $4$, it acts transitively on the projective points and lines in the quantum kernel, and moreover, if we consider the stabilizer of 
any point $v$ of $\mQ$ in the unitary group, it acts transitively on the points of $\mQ$ which are not orthogonal to $v$. For any division ring $D$ (with involution) and 
any dimension, similar transitivity properties are known. One way of understanding this action on $\mQ$ is through the theory of ``BN-pairs,'' see for instance \cite{Tits} and \cite{BueCoh} for the 
general case of division rings, and \cite{FGQ2} and \cite{SFGQ} for the modal case of finite fields in low dimensions (in which case $\mQ$ is denoted by $\mH(d,q^2)$; see section \ref{h3q}). 

It goes without saying that such properties are highly suited for designing quantum codes. In AQT, for instance 
in quantum super dense-coding, the quantum gates operators are also used in setting up the coding scheme; see sections \ref{super} and \ref{codes} for more details. We also note that since the unitary group $\hU(V,\Big \langle \cdot,\cdot \Big\rangle)$ acts on $V \setminus \mQ$, it is not possible to transport elements of $V \setminus \mQ$ to $\mQ$ through unitary operators.

\medskip
\subsection{Inner--Polar dichotomy}

The Inner--Polar dichotomy is the following principle: \\

$\left\{
\begin{tabular}{p{.9\textwidth}}\ul{although we might partially lose the interpretation of probabilities if the Hermitian form of the GQT is not an inner
product, in general we gain extra structure because the quantum kernel --- the geometry of the vectors or projective points which are self-orthogonal --- has a nontrivial 
geometric structure.}\end{tabular}\right\}$\\

\medskip
The word ``partially'' is important here: as we will see in section \ref{Born}, in all GQTs of algebraically closed fields in characteristic $0$ such as $\C$, we {\em do} have a 
probabilistic interpretation. The theories in which there is a nontrivial quantum kernel, hence appear to contain {\em extra} information (due to the geometric structure of the quantum kernel). 

In the next section, we give one GQT over the complex numbers which enjoys this principle in a most interesting way.  

\subsection*{Question}

What is the physical interpretation of the quantum kernel?

\medskip
\section{Other GQTs over $\C$ with nontrivial quantum kernel}

Let $\ell \geq 4$ be a positive integer. Define the following $(\sigma,1)$-Hermitian form on $V(\ell + 1,\C)$ or $\PG(\ell,\C)$, where $\sigma$ is complex conjugation (which we will denote by an overbar below):

\begin{equation}
\Big\langle x, y \Big\rangle = -(\overline{x_0}y_0 + \overline{x_1}y_1) + \overline{x_2}y_2 + \ldots + \overline{x_{\ell}}y_{\ell}. 
\end{equation} 

This form is not an inner product, but a nontrivial quantum kernel arises for each $\ell$; usually it is denoted by $\mH(\ell,\C)$, and as a projective variety, it has 
Witt index $1$. This means it only contains points and lines as full linear subspaces. It is a generalized quadrangle for each $\ell$ (see \cite{POL}). 

This GQT behaves much like AQT (it is obviously ``complex-like'' in the terminology of below, see section \ref{complike}), and also obeys the Born rule | see section \ref{Born}. On the other hand, the geometry of $\mQ$ (and the action of the unitary group upon $\mQ$) yields additional information. \\

Even if one wants to work with classical AQT over $\C$, one could still use the quantum kernel as follows. Embed AQT in a larger GQT (see section \ref{ext} for more details) over for instance an algebraically closed field $k$ that strictly contains $\C$, and extend the $(\sigma,1)$-Hermitian form over $\C$ to $k$ in such a way that a nontrivial quantum kernel arises over $k$. Classical AQT cannot see this kernel, but we can use it nevertheless, for instance in coding schemes such as the one presented in the final section of this paper.

\medskip
\section{Complex-like Quantum Theories, extension of Quantum Theories, and the minimal model}

As $\mathbb{C}$ is an algebraically closed field, the algebraically closed fields amongst the division rings deserve separate interest
in GQT. The following theorem is especially handy in this discussion.

\begin{theorem}[Baer]
\label{Baer}
Let $k$ be an algebraically closed field. Then $k$ has  nontrivial involutory automorphisms if and only if $k$ has nontrivial automorphisms of finite order if and only if
its characteristic is $0$.
\end{theorem}

So, each algebraically closed field $k$ of characteristic $0$ has a nontrivial involution $\sigma$ in $\Aut(k)$, and hence any vector space over such a field has Hermitian forms, while vector spaces over fields such as $\overline{\mathbb{F}_p}$ with $p$ any prime, cannot, and hence only have symmetric $\sigma$-sesquilinear forms.

We focus on characteristic $0$ now.

\subsection{Complex-like Quantum Theories}
\label{complike}

Let $k$ be any algebraically closed field in characteristic $0$. 
By Theorem \ref{Baer}, we know that there is an involution $\gamma$ in $\Aut(k)^\times$. Now consider the set
\begin{equation}
k_{\gamma} := \{ \kappa \in k\ \vert\ \kappa^{\gamma} = \kappa \}. 
\end{equation}

One easily shows that $k_{\gamma}$, endowed with the addition and multiplication coming from $k$, is also a field. There is however more (see \cite{ArtSch}).

\begin{theorem}[$(\mathbb{C},\mathbb{R})$-Analogy]
\label{analogy}
Let $k$ be any algebraically closed field in characteristic $0$. Let $\gamma$ be an involution in $\Aut(k)^\times$. Then $-1$ 
is not a square in $k_{\gamma}$. Suppose $i \in k$ is such that $i^2 = -1$. Then $k = k_{\gamma} + i\cdot k_{\gamma}$
and $[k : k_{\gamma}] = 2$.  
\end{theorem}

So each element of $k$ has a unique representation as $a + bi$, with $a, b \in k_{\gamma}$ and $i$ a fixed solution of $x^2 = -1$. Fields which have index $2$ in their
algebraic closure are called {\em real-closed fields}, and can always be constructed as a $k_{\gamma}$ of some involution $\gamma$. Real-closed fields
share many properties with the reals $\mathbb{R}$: each such field is {\em elementarily equivalent} to the reals, which by definition means that \ul{it has the 
same first-order properties as the reals.}  We call a GQT {\em complex-like} if it is defined over an algebraically closed field $k$ with nontrivial involution $\gamma$, 
where the elements of $k$ are represented in Theorem \ref{analogy} with respect to the field $k_{\gamma}$. 

The analogy goes even further: once we have defined $k_{\sigma}$ as above, and we represent each element $x$ in $k$ as $x = u + iv$, it can be shown that the automorphism
$\sigma$ is given by
\begin{equation}
\sigma:\ k \mapsto k:\ u + iv \mapsto u - iv. 
\end{equation}

(This is easy to see: as $u, v \in k_{\sigma}$, they are fixed by $\sigma$, and as $-1 \in k_{\sigma}$, it is also fixed. So either $\sigma(i) = i$ or $\sigma(i) = -i$. Since 
$\sigma$ is not trivial, we have the second possibility, and this concludes the proof.) \\

\subsection{Extension of Quantum Theories}
\label{ext}

If we consider a GQT over a field $k$ in characteristic $0$,  the fundamental question arises if it is {\em embeddable}
in a complex-like theory. Here, the notion of ``embeddable'' is obvious: if $k$ comes with the involution $\gamma$, we ask for a field extension $\ell \Big/ k$, 
where $\ell$ is algebraically closed, and an involution $\overline{\gamma}$ of $\ell$, such that the restriction of $\overline{\gamma}$ to $k$ is $\gamma$. Since any GQT
is only dependent on the Hermitian matrix of the $(\sigma,1)$-Hermitian form with respect to a chosen basis (with suitable adaptation to the infinite-dimensional case), 
it is clear that if the aforementioned GQT comes with matrix $A$ over $k$, then the same matrix $A$ defines a $(\overline{\gamma},1)$-Hermitian form over $\ell$ which 
induces the initial form over $k$.   
So if we fix the dimension of the Hilbert space, then any GQT over $k$ (and with respect to $\gamma$) is part of the GQT over $\ell$ (with involution $\overline{\gamma}$). Unfortunately, such a general result cannot hold. Consider for instance the field $\mathbb{Q}(\sqrt{2})$ with involution $\gamma$ generated by 

\begin{equation}
\sqrt{2} \mapsto -\sqrt{2}. 
\end{equation}

Then in \cite{Schnor} Schnor shows that although it does extend to an automorphism of $\C$, every extension of $\gamma$ would have infinite order. We adapt his proof to our setting, as follows. Suppose $\ell \Big/ k$ is an algebraically closed field with involution $\sigma \ne \id$, such that $\sigma$ extends $\gamma$. Then $\ell_\sigma$ is a subfield of index $2$ in $\ell$, and $\ell_{\sigma}$ is real-closed. Let $i \in \ell \setminus \ell_{\sigma}$ be such that $i^2 = -1$ (noting that $-1 \in \ell_{\sigma}$). As $\ell = \ell_{\sigma}(i)$, it follows that 
$\sigma(i) =-i$. So $\mathbb{Q}(\sqrt{2}i) \subseteq \ell_{\sigma}$, as $\sigma(\sqrt{2}i) = \sigma(\sqrt{2})\sigma(i) = \sqrt{2}i$. In $\mathbb{Q}(\sqrt{2}i)$ we have the identity
\begin{equation}
-1 = -2 + 1 = (\sqrt{2}i)^2 + 1^2,
\end{equation} 
so $-1$ is a sum of squares in the real-closed field $\ell_{\sigma}$, contradiction. \\

However, as Schnor points out in \cite{Schnor}, the result is true if we suppose $k$ to be algebraically closed to begin with.

\begin{theorem}[Embedding Theorem of Quantum Theories]
\label{emb}
Any GQT over an algebraically closed field $k$ with involution $\gamma$ is embeddable in a GQT over $\ell$, where $\ell$ is any algebraically closed field extension 
of $k$.
\end{theorem}

\medskip
So even if one prefers to keep working with AQT over $\C$, it is a fundamental fact that \ul{AQT is embedded in a universe of GQTs which extend AQT, in many of which 
quantum kernels arise which cannot be seen in AQT, but which are virtually there.} They should supply an extremely rich and powerful source for new tools in, e.g., Quantum Information Theory.

\medskip
\subsection{The minimal model: $\overline{\mathbb{Q}}$}

The following result is a well-known standard result on sizes of algebraically closed fields. 

\begin{theorem}
\label{count}
Let $k$ be any field. If $k$ is not finite,  its algebraic closure $\overline{k}$ has the same size as $k$; if $k$ is finite, $\overline{k}$ is countable.
\end{theorem}

Each field of characteristic $0$ contains the characteristic $0$ prime field: the field of rationals $\mathbb{Q}$. So each algebraically closed 
field $k$ in characteristic $0$ contains the algebraically closed field $\overline{\mathbb{Q}}$: \begin{equation}
\overline{\mathbb{Q}} \subseteq k.
\end{equation}

By Theorem \ref{count}, $\overline{\mathbb{Q}}$ is countable, and hence also minimal in size with respect to being algebraically closed.

Each GQT over $\overline{\mathbb{Q}}$ can hence be seen as a minimal model for other GQTs in characteristic $0$. By the Embedding Theorem \ref{emb}, any minimal GQT can be 
embedded in a GQT over any other given algebraically closed field in characteristic $0$. In the other direction, if $k$ is algebraically closed in characteristic $0$, and we consider a GQT over $k$ with involution $\sigma$, $\sigma$ fixes the prime field $\mathbb{Q}$, so also $\overline{\mathbb{Q}}$ (the induced action might be trivial, of course, and then the induced quantum theory over $\overline{\mathbb{Q}}$ is orthogonal/symmetric).

\medskip
\section{All the general Born identities}
\label{Born}

As we have seen in the previous  section, each GQT over an algebraically closed field $k$ in characteristic $0$ is complex-like, and all such theories essentially contain the minimal model. In this section we note that even more, all such theories enjoy precisely \ul{the same Born rule} as AQT does.  

So suppose we consider such a GQT, and let $\sigma$ be the corresponding involution. We represent each element in the algebraically closed field $k$ as $a + ib$, with 
$i^2 = -1$ and $a, b$ in the fixed field $k_{\sigma}$. Now let a quantum system be described by the wave function $\vert \psi \rangle$, and consider an observable which 
corresponds to the Hermitian operator $A$, with orthonormal base of eigenvectors $\{ \vert \phi_1 \rangle, \ldots, \vert \phi_d \rangle \}$. To each $\vert \phi_i \rangle$ is associated an eigenvalue $\lambda_i$ of $A$. Then the wave function $\vert \psi \rangle$ assigns a probability amplitude $\langle \phi_i \vert \psi \rangle \in k$ to the possible outcome measurement $\lambda_i$. If $\langle \phi_i \vert \psi \rangle = u + iv$ with $u, v \in k_{\sigma}$, then we define the {\em probability} of the measurement $\lambda_i$ to be 

\begin{equation}
\label{eqBorn}
\Big\vert \langle \phi_i \vert \psi \rangle \Big\vert^2 \ =\ u^2 + v^2. 
\end{equation}

The latter quantity is an element of $k_{\sigma}$. Now since $k_{\sigma}$ is real-closed, it is well known that squaring naturally induces a total order $\leq$ on $k_{\sigma}$, in which the 
set of all squares precisely is the set of positive numbers relative to $\leq$. In this total order, we say that $a < b$ if and only if we can find some square $c^2$ such that $a + c^2 = b$.
So each quantity coming from the general Born rule (\ref{eqBorn}) is a positive number, and two such quantities can be compared in the naturally defined total order. In the particular case of

\begin{equation}
k = \C, \ \sigma = \mbox{complex conjugation},\ k_{\sigma} = \mathbb{R},
\end{equation}
the induced total order is the standard one in $\mathbb{R}$, and we then obtain the classical Born rule. \\

We cannot have the same interpretation in arbitrary other division rings with involution. For instance, in the case of finite fields $\F_{q^2}$ (section \ref{secMQT}), we can indeed 
define, once we have chosen $\kappa$ as in that section, $\vert \langle \phi_i \vert \psi \rangle \vert^2$ to be $a^2 + b^2 \in \F_q$, but finite fields do not admit 
a total order, so we lose our classical interpretation of probabilities. On the other hand, we might interpret the quantity $a^2 + b^2$ as some kind of probability by which 
we can distinguish between different probabilities, but not say which one is the more probable than the other, unless one is $0$. The extreme case is 
$\F_q = \F_2$, where we only have the probabilities $0$ and $1$.

\medskip
\section{No cloning in GQT}
\label{noclone}

In this section we will obtain a no-cloning result, similarly as in AQT and MQT | see \cite{MQT}. We will consider  a GQT 
$\mQ = \Big((k^n,+,\cdot),\Big\langle \cdot,\cdot\Big\rangle\Big)$ over any division ring $k$; the result, which generalizes the results of \cite{MQT,WZclone}, is proven in a slightly different manner, since $k$ is not necessarily commutative. \\

Let $\mH_A$ and $\mH_B$ be copies of a Hilbert space $\mH$, and consider two states $\vert \phi \rangle_A$ and $\vert \psi \rangle_A$ in $\mH_A$. Suppose $U$ is a cloning unitary operator so that, for any $\alpha, \beta \in k$, we have
\begin{equation}
U\Big[(\alpha\vert \phi \rangle_A + \beta \vert \psi \rangle_A)\otimes \vert e\rangle_B \Big] = (\alpha\vert \phi \rangle_A + \beta \vert \psi \rangle_A)\otimes(\alpha\vert \phi \rangle_B + \beta \vert \psi \rangle_B),
\end{equation}
where $\vert e \rangle_B$ is an unknown blanco state (and $\otimes$ is the usual matrix tensor product). 
Using linearity of the operator on the left-hand side, and writing out the right-hand side, we obtain that
\begin{equation}
\label{meq}
\alpha\vert \phi \rangle_A \otimes \vert \phi \rangle_B + \beta\vert \psi \rangle_B \otimes \vert \psi \rangle_B =  
(\alpha \vert \phi \rangle_A)\otimes(\alpha \vert \phi \rangle_B) + (\alpha \vert \phi \rangle_A)\otimes(\beta \vert \psi \rangle_B) + (\beta \vert \psi \rangle_A)\otimes(\alpha \vert \phi \rangle_B) + (\beta \vert \psi \rangle_A)\otimes(\beta \vert \psi \rangle_B).
\end{equation}

First note that if $k$ is commutative and
$\phi_A$ and $\psi_A$ would happen to be orthogonal, then  (\ref{meq}) leads to the conditions $\alpha^2 = \alpha$ and $\beta^2 = \beta$ (for all $\alpha, \beta \in k$), and this only is true when $k$ is the finite field $\mathbb{F}_2$. We proceed with the general case. \\

Let $\alpha = 1 = \beta$; then we get that
\begin{equation}
\label{meq2}
\vert \phi \rangle_A \otimes \vert \psi \rangle_B + \vert \psi \rangle_A\otimes \vert \phi \rangle_B = 0.
\end{equation}

Put $\vert \phi \rangle_A = (a_1 \ldots  a_n)^T$ and $\vert \psi \rangle_A = (b_1 \ldots b_n)^T$. Then (\ref{meq2}) implies that 
\begin{equation}
\label{meq3}
a_ib_j = -b_ia_j\ \ \ \forall i, j \in \{ 1,\ldots,n\}.
\end{equation}


Suppose w.l.o.g. that $a_1 \ne 0$. Then $(a_1^{-1}(-b_1))a_j = b_j$ for all $j = 1,\ldots,n$. So \ul{either $b_1 = 0$, and then $\vert \psi \rangle_A$ is the zero-vector, or $b_1 \ne 0$ and $\vert \phi \rangle_A$ and $\vert \psi \rangle_A$ are members of the same ray} (when we multiply by the left). (Note that in a division ring, there are no zero divisors.)

\subsection{Rays} 

Suppose $k$ is a field, and suppose $\vert \psi \rangle_A$ and $\vert \phi \rangle_A$ \ul{are in the same ray}. Write, for a fixed $\rho \in k^\times$, and using the same notation as in the previous paragraphs, 
\begin{equation}
\rho a_i = b_i \ \ \forall i. 
\end{equation}

Then (\ref{meq3}) implies that 
\begin{equation}
a_ib_j + b_ia_j  = 2\rho a_ia_j = 0\ \ \forall i, j.
\end{equation}

So either $2 = 0$, and $k$ has characteristic $2$, or $a_ia_j = 0$ for all $i, j$. In the latter case, we may take $i = j$, so $\vert \phi \rangle_A$ is the zero vector.
(If $k$ has characteristic $2$, and $\vert \psi \rangle_A \ne 0$, $\vert \phi \rangle_A \ne 0$ are cloned, and $\vert \psi \rangle_A + \vert \phi \rangle_A$ as well, 
$\vert \phi \rangle_A$ is in the same ray as $\vert \psi \rangle_A$.)
\\

Suppose $k$ is a division ring which is not a field, and suppose $\vert \psi \rangle_A$ and $\vert \phi \rangle_A$ \ul{are in the same ray}; let $\rho$ be as in the previous section.

Then for all $i, j$ we obtain 
\begin{equation}
\label{meq4}
a_i\rho a_j = -\rho a_j a_i.
\end{equation}

For $i = j$, we obtain that $a_i\rho = -\rho a_i$ for all $i$. Plugging the latter in (\ref{meq4}), we obtain 
\begin{equation}
\label{comm}
\Big[a_i,a_j \Big] = 0 \ \ \forall i, j. 
\end{equation}

Whence for division rings, \ul{nontrivial subsets of rays can be cloned.} In particular, by the above, we have that if $\vert \psi \rangle_A$, $\vert \phi \rangle_A$, and $\vert \psi \rangle_A + \vert \phi \rangle_A$ are cloned, then $\vert \psi \rangle_A = 0$ or $\rho\vert \psi \rangle_A = \vert \phi \rangle_A$ for some $\rho \ne 0$, and (\ref{comm}) holds; $\rho$ now can 
be chosen arbitrarily. Note that $\vert \psi \rangle_A + \vert \phi \rangle_A$ is of the form $(1 + \rho)\vert \psi \rangle_A$. 

Once one passes to projective space, rays become points.




\medskip{\rm
\begin{remark}[Permutation cloning in GQT and AQT]
Let $k$ a division ring with involution, consider a GQT $\mQ = \Big((k^n,+,\cdot),\Big\langle \cdot,\cdot\Big\rangle\Big)$  over $k$.
Let $S$ be any subset of $k^n$, and let $\vert e \rangle$ be a blank state. We say that a unitary operator $U$ {\em permutation copies} $S$
if 
\begin{equation}
U: k^n \otimes k^n  \mapsto k^n \otimes k^n:\ \   \vert \psi \rangle\otimes  \vert e   \rangle  \mapsto U(\vert \psi \rangle\otimes \vert e \rangle)
\end{equation}
induces a bijection between the set $S \otimes \vert e \rangle:= \{ \vert \psi \rangle \otimes\vert e  \rangle\ \Big\vert\ \vert \psi \rangle \in S\}$ and the set 
$\mC(S) := \{ \vert \varphi \rangle \otimes \vert \varphi \rangle\ \Big\vert\ \vert \varphi \rangle \in S\}$. Clearly, if $U$ permutation copies $S$, it induces a permutation on the elements of $S$ in a natural way, and we say that it is the {\em associated permutation}.

\subsection*{Examples, special cases}

\begin{itemize}
\item
Let $S = k^n$. As $U$ is unitary, it is  nonsingular, so permutation copies $S$. 
\item
Put $S = \{ \vert \psi \rangle \}$; then $U$ clones $\vert \psi \rangle$. 
\item
Suppose $U$ is a unitary operator which clones each element of some set $S$; then $U$ permutation clones $S$ with associated permutation
$\id$ (the trivial permutation). 
\end{itemize}
\end{remark}
}

\medskip
\section{No quantum deleting in GQT}
\label{nodelete}

In a similar fashion as for cloning, we now obtain a no-deleting result, as in AQT and MQT | see \cite{MQT,nodelete}. We will again consider GQT over any division ring $k$; we will give a short treatment that \ul{only uses the invertibility of the deleting operator.} 

Let $\mH_A$ and $\mH_B$ be copies of a general Hilbert space $\mH$, and consider two states $\vert \phi_A \rangle$ and $\vert \psi_A \rangle$ in $\mH_A$, and copies $\vert \phi_B \rangle$ and $\vert \psi_B \rangle$ in $\mH_B$. Suppose $U$ is an invertible operator (contained in $\mathbf{GL}_{d^2}(k)$) so that
\begin{equation}
U:
\begin{cases}
\vert \psi \rangle_A \otimes \vert \psi \rangle_B &\mapsto\ \vert \psi \rangle_A \otimes \vert e \rangle_B\\
\vert \phi \rangle_A \otimes \vert \phi \rangle_B &\mapsto\ \vert \phi \rangle_A \otimes \vert e \rangle_B,
\end{cases}
\end{equation}
where $\vert e \rangle_B$ is some unknown blank state in $\mH_B$. So $\alpha\vert \psi\rangle_A \otimes \vert \psi\rangle_B + \beta\vert \phi\rangle_A \otimes \vert \phi\rangle_B$ is mapped by $U$ to $\alpha\vert \psi\rangle_A \otimes \vert e \rangle_B + \beta\vert \phi\rangle_A \otimes \vert e \rangle_B$.

Since $U$ is invertible and linear, we have that for 
every $\alpha, \beta \in k$, that
\begin{equation}
U^{-1}:
\begin{cases}
(\alpha\vert \psi\rangle_A + \beta\vert \phi\rangle_A)\otimes\vert e \rangle_B &\mapsto\ (\alpha\vert \psi\rangle_A + \beta\vert \phi\rangle_A)\otimes(\alpha\vert \psi\rangle_B + \beta\vert \phi\rangle_B)\\
&= \alpha\vert \psi\rangle_A \otimes \alpha\vert \psi\rangle_B +  \alpha\vert \psi\rangle_A \otimes \beta\vert \phi\rangle_B + \beta\vert \phi\rangle_A \otimes \alpha\vert \psi\rangle_B + \beta\vert \phi\rangle_A \otimes \beta\vert \phi\rangle_B.
\end{cases}
\end{equation}

It follows that 
\begin{equation}
\alpha\vert \psi\rangle_A \otimes \vert \psi\rangle_B + \beta\vert \phi\rangle_A \otimes \vert \phi\rangle_B = \alpha\vert \psi\rangle_A \otimes \alpha\vert \psi\rangle_B +  \alpha\vert \psi\rangle_A \otimes \beta\vert \phi\rangle_B + \beta\vert \phi\rangle_A \otimes \alpha\vert \psi\rangle_B + \beta\vert \phi\rangle_A \otimes \beta\vert \phi\rangle_B.
\end{equation}

In particular, substituting $\alpha = \beta = 1$, we get
\begin{equation}
0 = \vert \psi\rangle_A \otimes \vert \phi\rangle_B + \vert \phi\rangle_A \otimes \vert \psi\rangle_B.
\end{equation}

The detailed analysis of this equation has been done in \S \ref{noclone}.


\medskip
\section{Quantum teleportation in GQT}

As in Bennet, Brassard, Cr\'{e}peau, Josza and Wootters \cite{QT}, one can show that quantum teleportation works in every GQT in much the same way as in \cite{QT}.
By quantum teleportation, we mean that Alice wants to send the information about some unknown state $\vert \phi \rangle$ to Bob, without sending the particle 
itself; she lets it interact with some other system (the ``ancilla'') in such a way that the particle then is in a standard state and the ancilla is in an unkown state 
containing all the information about $\vert \phi \rangle$. Now Alice sends Bob that ancilla, and Bob recovers a replica of $\vert \phi \rangle$ using unitary transformations. 
As in most papers on quantum teleportation, we will work with (variations of) Einstein-Podolsky-Rosen (EPR) states \cite{EPR} to perform the teleportation. \\

There is a catch though: \ul{if we have a division ring of characteristic $2$, the method fails}. We describe another way to handle this class of GQTs, which also works for the others. \\

Let us first mention that the {\em characteristic} of a division ring, is the smallest positive integer $n > 0$ such that $1 + 1 + \cdots + 1$ ($n$ times) equals $0$. If such an $n$ does not exist, by definition the characteristic is $0$. It is a basic fact that the characteristic of a division ring is always $0$ or a prime number. \\

Let the unknown state be the qubit $\vert \phi \rangle = \alpha\vert 0 \rangle + \beta\vert 1 \rangle$, relative to the computational base $\{ \vert 0 \rangle, \vert 1 \rangle \}$. 
We put $\vert 0 \rangle = (1, 0)$ and $\vert 1 \rangle = (0, 1)$.\\

$\circ$\quad
Define the Bell state $\mB$ as 
\begin{equation}
\mB :=  \vert 00 \rangle + \vert 1 1 \rangle.
\end{equation}
Here, we use the notation $\vert a b \rangle = \vert a \rangle \otimes \vert b \rangle$. We omit a factor $\sqrt{1/2}$, since otherwise we need to consider division rings in which this expression is well defined. \\

$\circ$\quad
Distribute $\mB$ to Alice and Bob; Alice and Bob each get two qubits, so that we write $\mB$ as 
\begin{equation}
\mB :=  \vert 0_A0_B \rangle + \vert 1_A 1_B \rangle.
\end{equation}

Now we describe the three qubit state of our system as
\begin{equation}
\label{eqqu}
\vert \phi_A \rangle \otimes \mB = \alpha\Big(\vert 0_A 0_A 0_B\rangle + \vert 0_A 1_A 1_B\rangle \Big) + \beta\Big( \vert1_A0_A0_B\rangle + \vert 1_A1_A1_B\rangle \Big).
\end{equation}

It is important to remark that this expression is indeed correct: each of the considered Kronecker products only contain $0$s and $1$s, and these commute
with any element of the division ring $D$. \\

$\circ$\quad
Now we introduce the states
\begin{equation}
\label{eqstate}
\begin{cases}
\vert \phi^+ \rangle = \vert 00 \rangle + \vert 11 \rangle\\
\vert \phi^- \rangle = \vert 00 \rangle - \vert 11 \rangle\\
\vert \psi^+ \rangle = \vert 01 \rangle + \vert 10 \rangle\\
\vert \psi^- \rangle = \vert 01 \rangle - \vert 10 \rangle.
\end{cases}
\end{equation}

Notice that if the characteristic of $D$ is $2$, $\vert \phi^+ \rangle = \vert \phi^- \rangle$ and $\vert \psi^+ \rangle = \vert \psi^- \rangle$!
If the characteristic is not $2$, then we can write 
 \begin{equation}
 \label{eq43}
\begin{cases}
\vert 00 \rangle = \frac{1}{2}\Big(\vert \phi^+ \rangle + \vert \phi^- \rangle\Big)\\ 
\vert 11 \rangle = \frac{1}{2}\Big(\vert \phi^+ \rangle - \vert \phi^- \rangle\Big)\\ 
\vert 01 \rangle = \frac{1}{2}\Big(\vert \psi^+ \rangle + \vert \psi^- \rangle\Big)\\ 
\vert 10 \rangle = \frac{1}{2}\Big(\vert \psi^+ \rangle - \vert \psi^- \rangle\Big).
\end{cases}
\end{equation}

Since $2 \ne 0$, $\frac{1}{2}$ is the well-defined and unique multiplicative inverse of $1/2$.

Substituting in (\ref{eqqu}) gives
\begin{equation}
\begin{aligned}
\vert \phi_A \rangle \otimes \mB &= \frac{1}{2}\Big( \vert \phi^+ \rangle \otimes (\alpha\vert 0_B\rangle + \beta \vert 1_B \rangle)\Big)
+ \frac{1}{2}\Big( \vert \phi^- \rangle \otimes (\alpha\vert 0_B\rangle - \beta \vert 1_B \rangle)\Big)
+ \frac{1}{2}\Big( \vert \psi^+ \rangle \otimes (\alpha\vert 1_B\rangle + \beta \vert 0_B \rangle)\Big) \\
&+ \frac{1}{2}\Big( \vert \psi^- \rangle \otimes (\alpha\vert 1_B\rangle - \beta \vert 0_B \rangle)\Big).
\end{aligned}
\end{equation}

Note that $\frac{1}{2}$ commutes with all elements of $D$. \\

$\circ$\quad
Alice subsequently performs a measurement, and she obtains one of $\vert \phi^+ \rangle, \vert \phi^- \rangle, \vert \psi^+ \rangle, \vert \psi^- \rangle$. 
Depending on the outcome, Bob's qubit is in the state $\alpha\vert 0 \rangle + \beta\vert 1 \rangle, \alpha\vert 0 \rangle - \beta\vert 1 \rangle, \alpha\vert 1 \rangle + \beta\vert 0 \rangle, 
\alpha\vert 1 \rangle - \beta\vert 0 \rangle$. \\

$\circ$\quad
Now Alice classically sends two bits to Bob which describe the outcome of her measurement.\\ 

$\circ$\quad
Bob uses this information to perform a unitary operation (one of the quantum logic gates) which transforms his qubit in the required state. \\

\medskip
\section*{The characteristic $2$ case}

When the characteristic of $D$ is $2$, the $4$ Bell states collapse into $2$ Bell states, and they generate a $D$-plane (the ``Bell-plane'') instead of $D^4 = V(4,D)$. The identities in (\ref{eq43}) also fail to be true (even with the factor $1/2$).
So we need a slightly different approach in the remaining cases. \\


$\circ$\quad
Introduce a state $\widetilde{\mB}$ as 
\begin{equation}
\widetilde{\mB} :=  \vert 01 \rangle + \vert 1 0 \rangle.
\end{equation}

As above, Alice and Bob share $\mB$ and $\widetilde{\mB}$. \\

$\circ$\quad
We describe the  state of our system as
\begin{equation}
\vert \phi_A \rangle \otimes \mB + \mathbf{I}^-\Big(\vert \phi_A \rangle \otimes \widetilde{\mB}\Big).
\end{equation}

Here, $\mathbf{I}^-$ is the $(8 \times 8)$-matrix with $1$'s on the anti-diagonal, and otherwise zeroes (it is unitary). \\

$\circ$\quad
Introduce the states
\begin{equation}
\begin{cases}
\vert \phi^+ \rangle = \vert 00 \rangle + \vert 11 \rangle\\
\vert \psi^+ \rangle = \vert 01 \rangle + \vert 10 \rangle.\\
\end{cases}
\end{equation}

$\circ$\quad
Re-write $\vert \phi_A \rangle \otimes \mB + \mathbf{I}^-\Big(\vert \phi_A \rangle \otimes \widetilde{\mB}\Big)$ as 
\begin{equation}
\vert \phi_A \rangle \otimes \mB + \mathbf{I}^-\Big(\vert \phi_A \rangle \otimes \widetilde{\mB}\Big) = \vert \phi^+ \rangle \otimes \Big(\alpha\vert 0 \rangle + \beta\vert 1 \rangle\Big) + \vert \psi^+ \rangle \otimes \Big(\alpha\vert 1 \rangle + \beta\vert 0 \rangle\Big).
\end{equation}

$\circ$\quad
Now proceed as before. \\

Note that this procedure works for \ul{all division rings, including all fields and thus also $\C$}. As a code, it is less secure though. \\

\medskip
\section{Super-dense quantum-coding in GQT}
\label{super}

Due to the same properties as in the previous section, the classical super-dense quantum-coding scheme \cite{Bennet} works for all GQTs. We use the following notation:

\begin{equation}
X := \begin{bmatrix}
0 & 1\\
1 & 0
\end{bmatrix}, \  \ Z := \begin{bmatrix} 1 & 0 \\
0 & -1
\end{bmatrix}, \ \ 
\bI_2 := \begin{bmatrix}
1 & 0 \\
0 & 1
\end{bmatrix}.
\end{equation}

The general scheme goes as follows.\\

$\circ$\quad
Alice starts with a classical message of two bits: $00, 10, 01$ or $11$. \\

$\circ$\quad
Together with Bob she prepairs the state $\vert 0_A 0_B \rangle + \vert 1_A 1_B \rangle$ (where we use the notation as before). \\

$\circ$\quad
After applying one of the four unitary quantum gate  operations $\Big\{ X, Z, ZX, \bI_2 \Big\}$ to her part of the state, Alice transfers her new qubit, which is one of the four states from eq. (\ref{eqstate}), through a noiseless qubit channel, to Bob.
She applies the operators as follows:\\

$\left\|
\begin{tabular}{p{0.9\textwidth}}
\begin{itemize}
\item
If the classical message is $00$, she leave the state invariant.
\item
If it is $10$, she (matrix) multiplicates it on the left with $Z \otimes \bI_2$.
\item
If it is $01$, she multiplicates it on the left with $X \otimes \bI_2$.
\item
Finally, if the message is $11$, she multiplicates it on the left with $ZX \otimes \bI_2$. 
\end{itemize}
\end{tabular}\right.$\\

\medskip
$\circ$\quad
Bob performs a Bell measurement in the Bell base. \\

$\circ$\quad
Since the Bell base vectors are mutually orthogonal, one is singled out. \\

$\circ$\quad
Knowing the shared coding scheme of Alice, 
Bob reconstructs the unitary operation which was applied and as such deduces what the classical message was. 
{\em Example:} if Bob receives $\vert 1_A 0_B \rangle + \vert 0_A 1_B \rangle$, he knows she applied $X \otimes \bI_2$, 
and this corresponds to the classical message $01$.\\

\medskip
\section*{The characteristic $2$ case}

If the characteristic of $D$ is $2$, the set of quantum gates operators collapses into a set of only two operators
\begin{equation}
\Big\{ X, Z, ZX, \bI_2 \Big\} \ \ \mapsto\ \ \Big\{ X,\bI_2 \Big\}.
\end{equation}

 So the collapsed Bell basis now generates, as before, 
a ``Bell plane'' 
\begin{equation}
V(2,D)\ \cong\ \Big\langle \vert 00 \rangle + \vert 11\rangle, \vert 01 \rangle + \vert 10\rangle  \Big\rangle. 
\end{equation}

The same super-dense coding scheme carries over without change (with the collapsed Bell states). \\
 


\medskip
\section{Quantum coding schemes using the quantum kernel}
\label{codes}

In this section we combine the super-dense coding scheme, with geometric properties of the quantum kernel, if nontrivial, to produce new coding schemes. We concentrate on the GQT over $\F_{q^2}$ of section \ref{h3q} (but many many variations are possible). The quantum kernel then is $\mQ = \mH(3,q^2)$, and the projective space $\bP = \PG(3,q^2)$. 
The involution $\gamma$ is given by $\gamma: \F_{q^2} \mapsto \F_{q^2}: r \mapsto r^q$. If we choose suitable homogeneous coordinates, the points of $\mQ$
satisfy the equation
\begin{equation}
X_0^{q + 1} + X_1^{q + 1} + X_2^{q + 1} + X_3^{q + 1} = 0.
\end{equation}

In the scheme explained below, Alice and Bob agree on three lines $U,V,W$ in $\mQ$ which do not intersect mutually, and also on some element $\eta$ of the unitary group $\mathbf{PGU}_4(q^2)$ (or $\mathbf{GU}_4(q^2)$ | the result is of course the same).  Since we cannot transport elements of $V \setminus \mQ$ to $\mQ$ (cf. subsection \ref{autq}), we 
will use geometric methods to realize this transport. \\

$\circ$\quad
Alice considers an arbitrary state $\vert \psi \rangle$, in $V(4,q^2)$ which is not self-orthogonal,  and considers the corresponding projective point $x$ in $\bP \setminus \mQ$. \\

$\circ$\quad
Alice applies the map $\pi$ (cf. section \ref{}) to $x$, to obtain a projective plane $\pi(x)$ (the dimension of $\pi(x)$ is $3 - \mathrm{dim}(x)$). \\

$\circ$\quad
She intersects $\pi(x)$ with $\mQ$ to obtain a {\em Hermitian curve} $\mC$ (see \cite[\S 7.4]{Hirsch}) (this is the point set of a nonsingular Hermitian variety in $\PG(2,q^2)$; it can always be given 
the equation $Y_0^{q + 1} + Y_1^{q + 1} + Y_2^{q + 1} = 0$ after suitably changing coordinates). \\

$\circ$\quad
Any line of $\mQ$ has precisely one point in common with $\mC$ (see \cite[\S 3.4]{FGQ2}), so intersecting $U, V, W$ with $\mC$, we obtain three distinct points 
$u, v, w$. \\

$\circ$\quad
Then Alice applies $\eta$ to obtain three distinct points $u', v' ,w'$, which are still contained in $\mQ$ since the unitary group acts on $\mQ$. \\

$\circ$\quad
Alice sends the points to Bob through a super-dense coding scheme.\\

$\circ$\quad
Bob reconstructs $u', v', w'$.\\

$\circ$\quad
Bob applies $\eta^{-1}$ to obtain three points $u, v, w$. He then constructs the projective plane $\Pi$ generated by these points.\\

$\circ$\quad
Bob applies $\pi$ to $\Pi$ and obtains $x = \pi(\Pi)$ (since $\pi^2$ acts as the identity). \\

\newpage
\medskip
\section{Concluding remarks}

In this paper, we proposed a unified way to consider classical and modal Quantum Theories, as governed by vector spaces over division rings $D$, endowed with 
$(\sigma,1)$-Hermitian forms ($\sigma$ an involution of $D$). We have shown that many foundational results such as no-cloning and no-deleting results, quantum teleportation
and super-dense coding can be obtained in one unified way in all these theories. Extra attention has been drawn to Quantum Theories over finite fields, and over 
algebraically closed fields. In case of the latter fields, we have shown that in characteristic $0$, the Quantum Theories behave much like classical Quantum Theory over 
the complex numbers, and that a fundamental extension theorem of theories applies. This provides the new possibility to enlarge classical Quantum Theory 
to larger algebraically closed fields and division rings, in which a singular geometrical object | the ``quantum kernel'' | arises, which carries the geometry of a polar space, and on which the full unitary group of the Quantum Theory acts. Modulo extension of theories, quantum kernels are always virtually around, depending on how one extends the theory. 
In this way, one can also easily switch between Quantum Theories when the nature of a physical situation would require a different angle of reasoning. We have used the quantum kernel in a new coding scheme over finite fields which mixes quantum teleportation together with geometric properties of the quantum kernel. All Quantum Theories over algebraically closed fields in characteristic $0$ share one model | the minimal model | which is defined over the algebraic closure of $\mathbb{Q}$ (so which is a countable model).

\end{document}